\documentclass{elsart}

\usepackage{amssymb}
\usepackage{graphicx}

\begin{document}

\begin{frontmatter}

\title{Time resolved particle dynamics\\ in granular convection}

\author[unav]{J.M. Pastor},
\author[unav]{D. Maza},
\author[unav]{I. Zuriguel},
\author[unav]{A. Garcimart\'{\i}n \corauthref{cor1}}
\ead{angel@fisica.unav.es} \corauth[cor1]{Corresponding author}
\author[cpmoh]{J.-F. Boudet}

\address[unav]{Depto. de F\'{\i}sica y Mat. Apl., Facultad de Ciencias ,\\Universidad de Navarra, 31080 Pamplona, Spain}
\address[cpmoh]{CPMOH, Universit\'e de Bordeaux I, 33405 Talence Cedex, France}

\begin{abstract}
We present an experimental study of the movement of individual
particles in a layer of vertically shaken granular material.
High-speed imaging allows us to investigate the motion of beads
within one vibration period. This motion consists mainly of
vertical jumps, and a global ordered drift. The analysis of the
system movement as a whole reveals that the observed bifurcation
in the flight time is not adequately described by the
Inelastic Bouncing Ball Model. Near the bifurcation point,
friction plays and important role, and the branches of the
bifurcation do not diverge as the control parameter is
increased. We quantify the friction of the beads against the 
walls, showing
that this interaction is the underlying mechanism responsible for
the dynamics of the flow observed near the lateral wall.
\end{abstract}

\end{frontmatter}

\section{Introduction}

Granular convection is a patent example of how collective movement
of grains can give rise to an ordered yet complex behavior. As
soon as 1831, M. Faraday \cite{Faraday} reported a long range
flow developed in a vertically shaken granular layer.
This flow is called \emph{granular convection} because
of the likeness between it and the movement of a liquid layer
heated from below.
Although many works have dwelt on this topic, the origin of this
convective movement, and in particular the role of the lateral
walls or the boundaries, is not fully understood. In 1989, P.
Evesque and J. Rajchenbach \cite{evesque} published an article
where they showed experimentally that the threshold for collective
motions to appear corresponds to the acceleration of gravity $g$.
This is why the acceleration of the external driving is often
given in the form of an adimensional number $\Gamma=\frac{A \omega
^2}{g}$, where $A$ is the amplitude and $\omega$ is the frequency
of the forcing. They also described that a heap grows changing the
shape of the \emph{free surface} of the medium, as a consequence
of the grains circulating in a ``convective'' fashion.

Almost at the same time, C. Laroche {\it{et al.}} \cite{laroche1}
reported both the importance of interstitial air for the
deformation of the granular layer and the development of a
compactation front that splits the layer into two zones, a
\emph{solid} one and a \emph{liquid} one. The origin of
convection, according to these authors, would be directly related to the
air circulating among the grains. This effect determines the
rising of material at the center of the layer and a flow of grains
going down near the walls, which influence the material by
increasing its porosity with respect to the central zone.
Nevertheless, subsequent studies \cite{clément}\cite{durand} have
revealed that the walls can by themselves provide the driving
force for convection, at least for a two-dimensional geometry.
This point was finally demonstrated by the works of the Chicago
group  \cite{nagel}, who used NMR techniques to show that wall
friction does affect the velocity profile of the particles. It
should be noted that the shaking was conveyed in this case in the
form of short pulses, or ``taps'', separated by rest periods much
longer than the pulse itself. At the same time, enlightening ideas
were put forward, and tested numerically, setting the framework in
which to understand the collective behavior of granular matter.
Following an analogy with hydrodynamics, models were developed
that qualitatively predicted the long range ordering of a shaken
granular media, even though simplifications sometimes made them
unrealistic \cite{kadanoff} \cite{Bourzosky} \cite{Hayakawa}
\cite{rosa}.

Above the convective threshold, a granular layer can also undergo
a rich array of instabilities. In 1989 S. Douady \emph{et al.}
\cite{douady} showed that beyond a certain value of $\Gamma$ the
flight of the grains experiences a period doubling bifurcation, in
a way essentially equivalent to a solid body that is placed on a
vibrating plate \cite{holmes}. As a layer of granular material is
strongly dissipative, it can be considered perfectly inelastic,
and its behavior as a whole can be modeled by an inelastic ball on
a vibrating plate. This simple model, known as the Inelastic
Bouncing Ball Model (IBBM) has been discussed by several authors
\cite{pieranski}\cite{tufillaro}\cite{mehta} and successfully
used to describe the temporal dynamics of a shallow granular layer
(without convection) \cite{pancho1} as well as the dilation of a
thick granular layer \cite{vandoorn}. Moreover, as the system is
spatially extended, it can also undergo spatial instabilities
associated to the breaking of translational symmetries between
different zones of the layer, that can oscillate with different
phases \cite{pancho1}.

More recently, it has been shown how the convective velocity field
depends on the adimensional acceleration and other parameters,
such as frequency and the air pressure \cite{nos} \cite{squires},
and new theoretical models have been developed
\cite{behringer3}, \cite{merson}, \cite{rodrigo}.

In this article we present an experimental study of the motion
that a dense granular system develops when it is submitted to a
vibration in the same direction than gravity. By detecting the
time at which the layer collides with the shaking plate, flight
times of the granular layer as a whole are measured, and the
temporal bifurcations are described. At the same time, the
movement of the particles near the container walls has been
tracked with a high-speed recording system. By tracking the grains
within an oscillation period, the friction effects caused by walls
can be quantified and its influence on the global circulation is
assessed.

\section{Experimental set-up}

\begin{figure}[tb]
\begin{center}
\includegraphics[width=9cm]{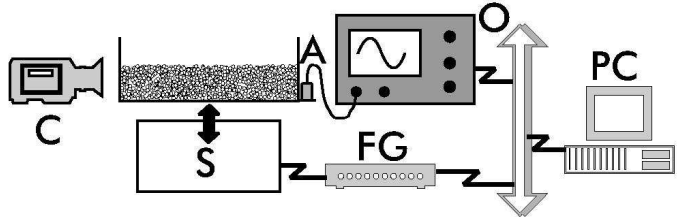}

\caption{Experimental set-up. A granular layer is vibrated by
means of a shaker (S), which is in turn controlled by a function
generator (FG). The acceleration is measured with an accelerometer
(A). An oscilloscope (O) is used to monitor the instantaneous 
acceleration. The movement of the particles adjacent to the wall is
recorded with a high-speed camera (C). The devices are
controlled from a PC.} 

\label{fig.setup}
\end{center}
\end{figure}

Convection can be observed with almost any granular material,
irrespective of shape, size or surface features. We have used
glass beads with a diameter of $0.5 \pm 0.1$ mm, but we checked
that sand gives the same qualitative results. The relative effect
of cohesive forces (such as humidity or static charges) is
important if the beads are much smaller than this. On the other
hand, it is desirable to have as many grains as possible, and this
particular size offers a good compromise. We put a big number of
beads (typically of the order of $10^4$) inside a cylindrical box
made of glass. The fact that both the beads and the box are of the
same material reduces the amount of electrical charges created by
friction. We also sprinkled the box with antistatic spray. The
diameter of the box is 52 mm and it is high enough to avoid grains
falling over when vibrated. This box is attached to a TiraVib
52100 magnetic shaker capable of delivering a sinusoidal
acceleration of up to 15 $g$ with a distortion smaller than 0.05
$g$. The shaker is commanded by a Stanford Research DS345 function
generator. The vibration is characterized with an accelerometer
attached to the box that has a sensitivity of $100\;mV/g$, whose
signal is picked by a Hewlett-Packard HP54510 digitizing
oscilloscope. Both the oscilloscope and the function generator are
connected to a PC. A sketch is provided in Fig.~\ref{fig.setup}.

The frequency of the external vibration $f$ was kept constant at
$f=110\;Hz$. We had previously found that the features of
convection do not change qualitatively with frequency \cite{nos}
provided that it is higher than 60 Hz. The acceleration was
therefore changed by regulating the vibration amplitude $A$. The
size of the granular layer is given in terms of the dimensionless
height $N=h/\phi$, where $h$ is the thickness of the layer and
$\phi$ the particle diameter.

We used a high-speed camera (Motionscope Redlake, model 1105-0003)
with a macro lens and a VCR to record the movement of the grains
at 1000, 2000 or 4000 frames per second. Under proper
illumination, each glass sphere will reflect a bright spot that
has been tracked with the following procedure. Once transferred to
the computer, the movie was split into individual frames. A
morphological image processing was performed on each frame to
obtain the centroid of one bright spot in the first recorded
frame. As the spheres budge less than one diameter from one frame
to the next, the position of the bright spot is easily
identifiable in the subsequent frame. Repeating the procedure for
all the recorded frames and by tracking several beads, a set of
grain positions versus time was obtained from each movie. Note
that only spheres adjacent to the walls are accessible with this
method, and we can only measure the velocity in the plane of the
wall. An alternative method that has also been used, yielding the
same results, is to calculate the correlation function between
consecutive frames. In this case, the averaged velocity of all the
beads in the frame is obtained.

\section{Motion of the center of mass}

\begin{figure}[tb]
\begin{center}
\includegraphics [width=14 cm]{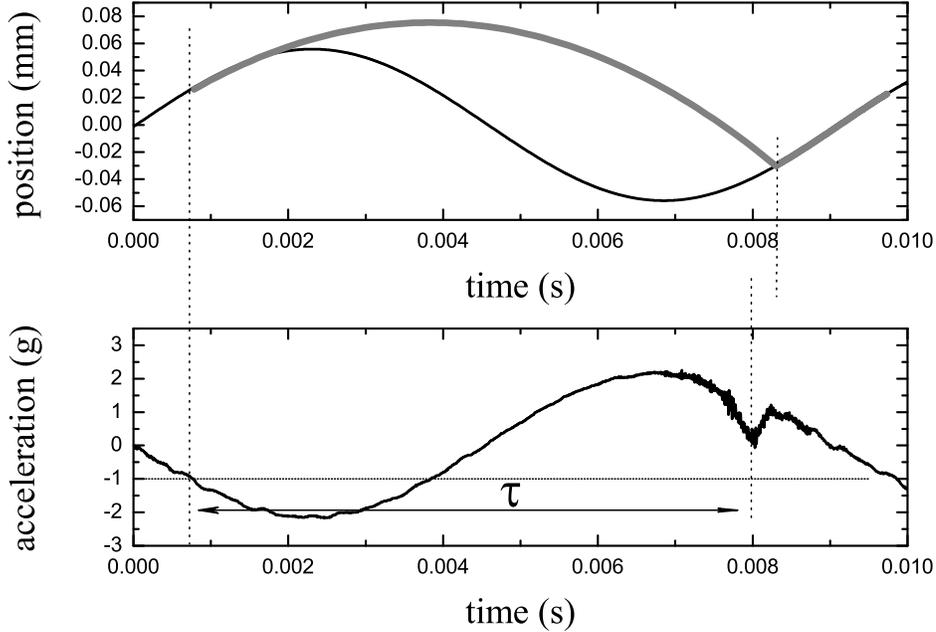}

\caption{ Parabolic flight predicted by the Inelastic Bouncing Ball
Model ({\it top}) and the acceleration measured by the
oscilloscope for $\Gamma=2.37$ and $N=33$ ({\it bottom}). The
collisions of the granular layer against the base of the container
are evident in the signal from the accelerometer. The value of the
acceleration equal to $-g$ is marked with a horizontal dotted
line; the granular layer gets loose at this coordinate.
Remarkably, the flight time measured in the experiment, $\tau$ 
suffers a phase delay with regard to the predicted by the IBBM.
Take off and collision are marked with vertical dotted lines.}

\label{fig.ibbm}
\end{center}
\end{figure}

Let us begin by describing the motion of the layer as a whole
without considering the movement of the individual grains. Under
this assumption, and considering the layer as a perfectly
inelastic solid, its center of mass will begin to fly when its
acceleration overcomes the gravity. From then on, the material
will perform a free flight, and will lose all its energy when it
collides with the plate. If the acceleration of the container is
at that moment smaller than $g$, as in Fig.~\ref{fig.ibbm}.a, the
layer gets stuck to the container base until it departs from the
base again when the acceleration exceeds the
gravity. The layer spends therefore a time $\tau$ in the flight
and a time $T - \tau$ (where $T$ is the period of the container
oscillation) stuck to the container base in each cycle.

The flight time $\tau$ grows with $\Gamma$ until it reaches the
value of oscillation period of the forcing $T$. If the granular
layer is considered as a point mass, this happens for
$\Gamma=\sqrt{1+\pi^{2}}$. At that point, the flight time
undergoes a saddle-node bifurcation with the stable branch
corresponding to a flight time $\tau=T$ \cite{holmes}. This lasts
until $\Gamma=\sqrt{4+\pi^{2}}$, where a period doubling
bifurcation takes place. Beyond that point, the particle can
either perform a \emph{long} flight (longer than $T$) or a
\emph{short} flight (shorter than $T$), depending on the container
acceleration at the time of the collision. As $\Gamma$ increases,
the long flight grows longer and the short flight shorter. Above
$\Gamma= \sqrt{1+4\pi^{2}}$ only the long flight survives, and
when it reaches the value $2T$ it bifurcates again.

\begin{figure}[tb]
\begin{center}
\includegraphics[width=14cm]{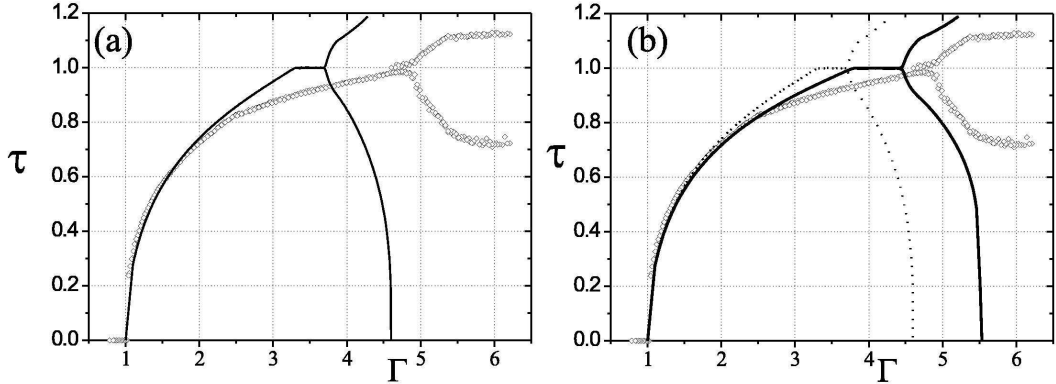}

\caption{(a) Dimensionless flight time $\tau$ of the granular
layer measured in the acceleration signal (\emph{symbols}). When
$\tau$ reaches $T$ the flight time undergoes a period doubling: a
long and a short flight are performed every two cycles. The line
indicates the values predicted by the IBBM. (b) The same data but
now the solid line is a numerical simulation including the air
effects (as in the Kroll analysis). This analysis improves
the fit somewhat. The dotted line is the same than the solid
one in (a). } 

\label{fig.bifurcacion}
\end{center}
\end{figure}

The validity of this model to reproduce the interaction of the
granular layer as a whole with the vibrating plate can be assessed
by comparing its predictions to the experimental measurement of
$\tau$ (see Fig.~\ref{fig.ibbm}.b). In order to do this, we have
taken the value predicted by the IBBM for the phase at the
beginning of the flight: $\phi_{i}=arcsin(1/\Gamma)$. There is no
way to obtain this value from the acceleration signal, as the take
off does not leave any trace on it. In principle this value is not
prone to be affected by the friction between the grains and the
container or the presence of interstitial gas, because at take off
the relative velocity between the granular layer and the vibrating
plate is zero. The collision time can be obtained experimentally
from the measured acceleration, as in Fig.~\ref{fig.ibbm}.b. The
flight times obtained in this way, subtracting the take-off times
from the collision times, are displayed in Fig.
~\ref{fig.bifurcacion}.a along with the bifurcation diagram
predicted by the IBBM. Clearly the model reproduces quite well the
flight times of the center of mass for $\Gamma \lesssim 2.7$, as
has already been demonstrated \cite{vandoorn}. Above this value,
the model is not valid anymore. The measured flight times are
shorter than those predicted, and the bifurcation point is at
$\Gamma=4.8\pm0.1$. Beyond that point, the branches grow but 
eventually they seem to saturate. Another remark is that the 
system bifurcates
directly from a monotonically growing solution to a period two
solution, without showing the saddle-node bifurcation predicted by
the model.

It is conceivable that one cause for this behaviour could be
associated to the effects of the interstitial gas on the granular
layer. The layer should then be considered as a porous medium
whose porosity changes as it detaches from the base. A pioneering
study of those dynamics has been done by Kroll \cite{kroll} and
refined by Gutman \cite{gutman}, who introduced air compressibility
and a coupling with the porosity of the medium.

Let us introduce the hypothesis of Kroll (which is easier than 
Gutman's to
perform)  in the numerical simulation of our problem. Considering
the inelastic ball as a porous piston interacting with the air in
the cell (a similar analysis has been recently reported for the
case of granular segregation \cite{colombia}), the numerical
analysis of the flight time suggests that air effects on the systems
should be measurable but ought not to change the dynamics
(see Fig.~\ref{fig.bifurcacion}.b),
\emph{i. e.} the saddle-node bifurcation is still present (its
range of stability is even increased) and the branches diverge as
the flight time approaches $2T$.

\begin{figure}[tb]
\begin{center}
\includegraphics[width=14 cm]{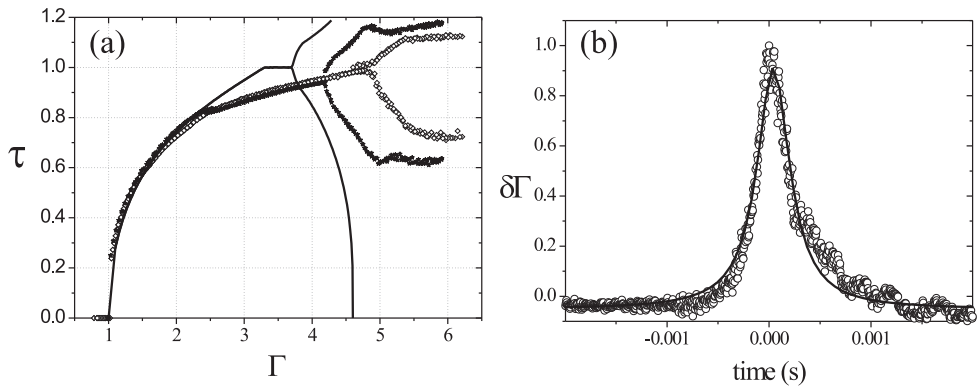}

\caption{Dimensionless flight time $\tau$ of the granular layer in
an evacuated container (\emph{filled symbols}). The agreement with
the IBBM is better than in air, but only for $\Gamma<2.3$. Above 
this value and up to the bifurcation, the dynamics of the system 
does not conform to the IBBM and coincides with the behaviour in 
the presence of air.
(b) The collision retrieved from the measured acceleration, 
normalized to the maximum height (for $\Gamma=3$). The collision 
between the layer and the vibrating plate takes place during
a finite time that can be measured from these data.}

\label{fig.vacio}
\end{center}
\end{figure}

In order to test the air influence on the flight time, we evacuated
the container to $10^{-2}Torr$ and we collected data for the same
range of $\Gamma$. Results are shown on Fig. ~\ref{fig.vacio}.a.
The agreement of measured data in vacuum with the IBBM is
excellent up to $\Gamma \sim 2.3$, better than in the presence of
air. Up to that value, flight times are a bit shorter if there is
interstitial air. Nevertheless, above $\Gamma \sim 2.3$ the measured
flight times do not fit to the IBBM even in vacuum. The
bifurcation point is noticeably changed, even though it is still
beyond the place predicted by the IBBM; the saddle-node
bifurcation is neither observed. Beyond the bifurcation, the
branches behave similarly in both cases (in vacuum and in air); a
remarkable feature in vacuum is a region where flight times are
bivaluated ($4.7<\Gamma<5$). It is interesting to note the
similarity of the branches between them and the likeness to the
branch before the bifurcation point; this will be analyzed
elsewhere.

From the comparison of both curves (see Fig. ~\ref{fig.vacio}.a),
the one in air and the one in an evacuated container, it is
evident that although air does affect granular convection, this is
manifested basically in the location of the bifurcation point,
that gets closer to the predicted by the IBBM. It is reasonable to
think that a higher vacuum would lead to an even better agreement.
But it still does not explain why the branches do not diverge and
why the saddle-node bifurcation goes unobserved.

Another feature that is lacking in the model is the
finite duration of the collision between the inelastic ball and
the plate. The extent of this time is a considerable portion of
the oscillation period $T$, as can be appreciated in  Fig.
~\ref{fig.vacio}.b. This implies that the velocity of the center
of mass at take off is not necessarily well defined. If the
collision lasts for some time, it is reasonable to think that
there is a delay that leads to a decrease in the initial velocity
of the center of mass, and therefore to shorter flight times. This
phenomenon is associated to the propagation of a shock wave front
\cite{bougie} that will be described elsewhere, and it has
significant consequences for $\Gamma>3$, when the duration of the
collision becomes similar to the time that the granular layer
spends stuck to the vibrating plate. For flight times shorter than
$80\%$ of the period, however, a finite
collision time should not affect the flight time and we should
search for another cause.

The key could be the interaction of the particles with the lateral
wall of the container. Thus, the wall would exert an effective
force on the inelastic ball larger than gravity that would affect
not only the initial phase of the flight \cite{durand} (its
effects on the initial velocity being negligible) but the
acceleration during the entire flight as well, resulting in an
effective force applied to the grains during the flight bigger
than gravity. Assuming that this force is independent of the
relative movement between the particles and the container wall, we
can estimate its value by comparing the measured flight times with
those predicted by the IBBM. We therefore introduce an effective
control parameter such as
$\Gamma_{eff}=\frac{A\omega^{2}}{g_{eff}}$ which depends on an
effective acceleration $g_{eff}$ whose value is $10.6\;m/s^2$. 
This is the value that must
be introduced in the IBBM in order to recover the measured flight
time. This will be described in detail in the next section, where
the movement of individual grains is dealt with.

\section{The motion of individual grains}

Till here we have described the motion of the granular layer as a
whole. But there is motion in the frame of reference of the layer:
the convective flow. Let us now study the movement of individual
particles. The convective motion --it has been described
previously \cite{nos},\cite{nagel} -- is much slower than the
vibration, so it could somehow be expected that the motion of
individual beads is a combination of flights similar to those of
the IBBM coupled with a slow drift.

In our experiment we have tracked the position of individual beads
near the lateral wall. The measurement have been performed near
the surface, where the downward velocity peaks.
The paths of the beads in the plane of the
wall do not divert much from the vertical: azimuthal velocities
are typically less than $10 \%$ of the vertical velocities. Note,
however, that there can be motion in the radial direction; this
component is not accessible in our experiment. Therefore we will
restrict in the following to the vertical direction except when
explicitly indicated.

\begin{figure}[tb]
\begin{center}
\includegraphics[width=14 cm]{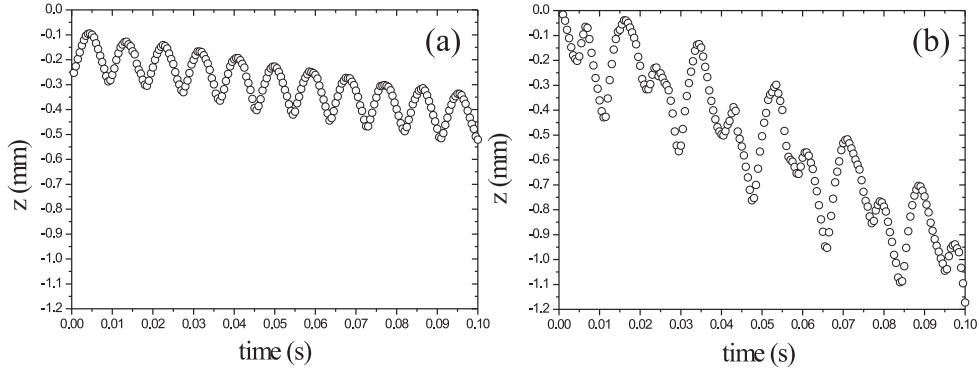}

\caption{The vertical position of a bead, tracked during about 10
cycles at 2000 samples per second. The container was being
vibrated at $\Gamma = 2.59 $ ({\it a}) and at $\Gamma = 5.81 $
({\it b}). The height of the layer was $N=100$ in both cases. The
origin of distances is arbitrary. Note the increasing in the drift
velocity when $\Gamma$ grows.}

\label{fig.saltos}
\end{center}
\end{figure}

\begin{figure}[tp]
\begin{center}
\includegraphics[width=8cm]{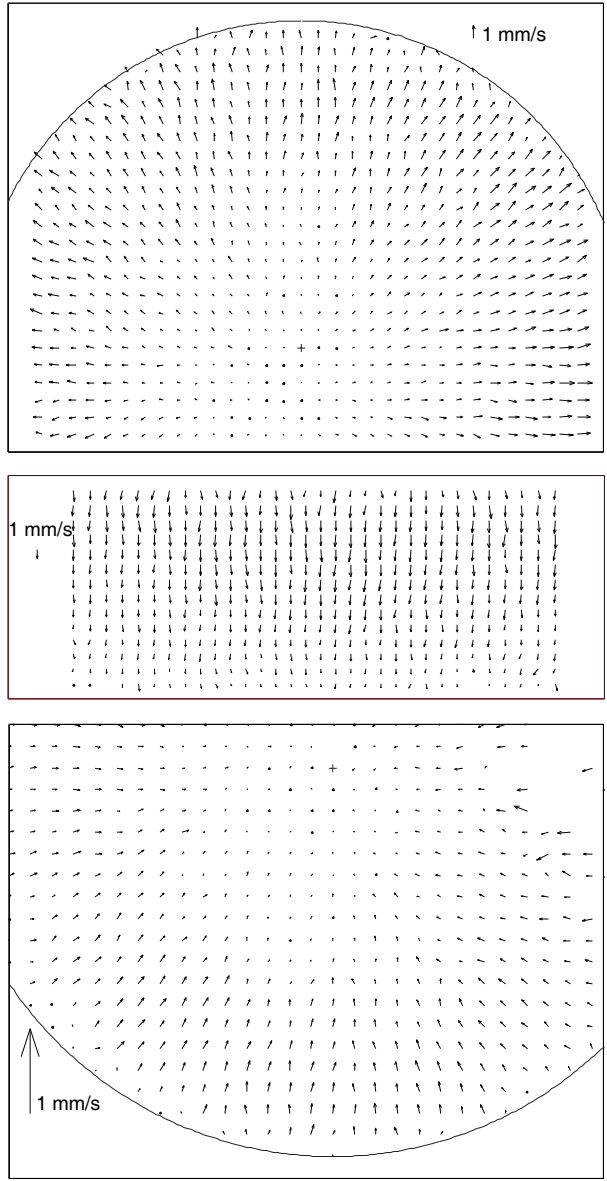}

\caption{ The convective velocity field at the top of the layer,
near the lateral wall and at the bottom of the layer ({\it from
top to bottom; only part of the layer is represented.}) Note the
different scales. This figure corresponds to $N=33$ and
$\Gamma=1.90$. The velocity field has been obtained by particle
tracking at a small sampling rate (25 frames per second)
effectively filtering out the rapid movement at the excitation
frequency. The small crosses in the top and bottom plots mark the
center of the container. } 

\label{fig.campo}
\end{center}
\end{figure}

The vertical position of a single bead at $\Gamma$ below and above
the period doubling point is plot on Fig.~\ref{fig.saltos}. The
beads roughly follow the same sort of movement described by the
IBBM, with a conspicuous difference: there is a distinct drift
downwards. We can consider the motion as consisting of two
components: a fast one (the jumps at frequency $f$) and a slow
one, which is the drift. The velocity
of the latter (the convective motion) is about an order of
magnitude smaller than the peak value of the former. Below
$\Gamma\simeq6$, we have observed that this configuration always
forms a toroidal convective roll: the beads go down near the wall
and they rise near the center of the container (see
Fig.~\ref{fig.campo}).

\section{Friction against the walls}

\begin{figure}[tb]
\begin{center}
\includegraphics[width=14cm]{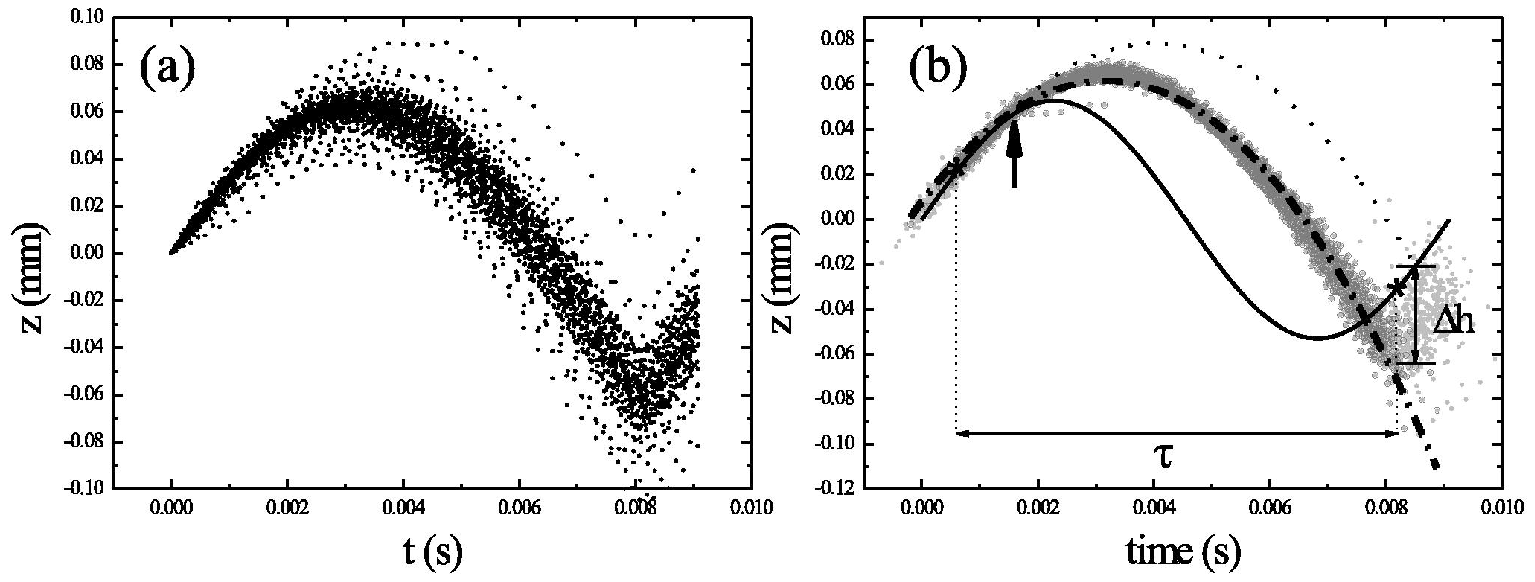}

\caption{(a) The trajectories of many grain flights are plotted
together. The origins of the positions are the places where
velocity changes sign. The data corresponds to $\Gamma=2.5$,
a value of the forcing where flight times for
an evacuated container and a container with air are the
same. It is difficult to fit these data because of their
dispersion.
\newline (b) The same data, but now the maximum height
has been chosen as the common reference. In order to relate 
them to the movement of the vibrating plate, the phase at the
moment of collision has been adjusted to the experimental
data (this point is marked with a star in the plot). From
this phase, the experimental points have been displaced so
their trajectories are tangent to the base at the beginning
of the flight.
The dotted line corresponds to the flight as predicted by the
IBBM. Particles do reach a lower height, and they finally
go down further than expected from a ``free'' flight. The
difference $\Delta h$ divided by the oscillation period gives
the velocity of the convective flow. The dashed line is a
quadratic fit.
Although the fit could be deemed god, the initial phase for 
the flight as obtained from the fit (arrow) does not coincide 
with $\Gamma=g$. }

\label{fig.vuelos}
\end{center}
\end{figure}

We are now able to discuss the origin of the downward movements
near the lateral wall. The motion of a single particle has been
shown to consist of a ``fast'' component (the jumps at the driving
frequency $f$ or $f/2$) and a ``slow'' component (the drift giving
rise to convection). Clearly, there must be some mechanism
imposing a net shear on the grains in order to induce the
flow depicted in Fig.~\ref{fig.campo}. 
With the aim of investigating this subject, we took a closer 
look at the trajectories of the particles during each cycle.

A large number of trajectories of \emph{single grain flights} 
recorded during one cycle, such as those displayed
in Fig.~\ref{fig.saltos}, is shown in Fig.~\ref{fig.vuelos}.a
The common origin has been chosen as the moment when
the particles
collide with the base. Obviously, the measured positions of individual
beads are too noisy to obtain clean paths. This is mainly due to
the rearrangements of grains during the flight and to the fact 
that beads rotate. In order to
regroup all the trajectories, the origin for all the paths has
been arbitrarily chosen at the maximum of the flight. Near this
point, almost all of the trajectories should have the same dynamics.
We implemented an algorithm to find the maximum that
chooses seven points around this zone an fits the trajectory to a
parabolic flight, and then regroups all the trajectories using 
the calculated maximum as the reference. Using this method, 
the temporal coordinate
where the particle touches the base has a lower dispersion. We use
the mean value of these coordinates and the corresponding time
determined from the oscilloscope to adjust the phase between the
vibrating plate and the flights of the grains.

In figure ~\ref{fig.vuelos}.b we show the regrouped trajectories
compared with the trajectory predicted by the IBBM. (We have
displaced vertically the measured trajectories so that they are
tangent to the plate oscillation). It seems clear that although
the particles begin the flight for a phase almost equal to the one
predicted by the IBBM, they finish the flight well below the
position were they would land after a free flight. This
difference, averaged for an oscillation period, is just the
velocity of the slow drift giving rise to the convective flow
\cite{nos}.

From the comparison between the experimental data and the
trajectory predicted by the IBBM it is clear that an acceleration
larger than gravity is acting on the particles (as can be seen in
Fig. ~\ref{fig.vuelos}. b, this can be easily deduced because the
heights reached in the flights, with the same initial condition,
are rather different in the measured data and the IBBM).
If this acceleration is constant during the flight time, the path
of the particles in a space-time diagram should fit to a parabolic
trajectory. We have fit a parabola, leaving all the parameters
free except the maximum, which we have taken as a common origin
for all the grains. This fit is shown in Fig. ~\ref{fig.vuelos}.
b. The fit seems acceptable, and the effective acceleration during the
flight would be $g_{f}=10.63\pm0.04\;m/s^2$. This value matches the 
$g_{eff}$ estimated from the flight times. Although the data are 
fitted reasonably well, the match is not completely
satisfying. As stated above, the position of the maximum is fixed
and then the acceleration and the initial velocity are free
parameters of the fit. The value obtained for the initial velocity
does not agree with the one predicted by the IBBM, and differs
from the one that can be extracted from the data. 

\begin{figure}[tb]
\begin{center}
\includegraphics[width=14cm]{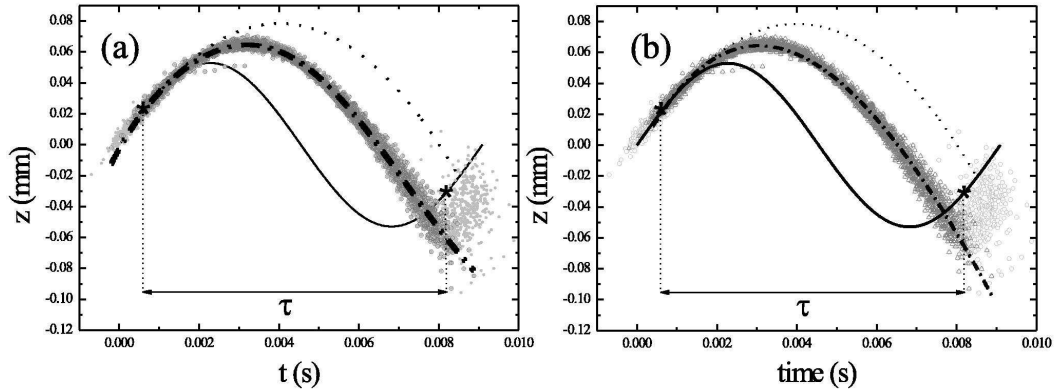}

\caption{(a) The same data than in Fig. \ref{fig.vuelos} but with
a viscous dissipation that depends on the velocity of the
grains included in the simulation. Now the
fit (dot-dashed line) is better, and the initial phase for the flight is
correctly predicted. The value obtained for the effective 
acceleration is now $g_{\nu}=12.77\pm0.07\;m/s^2$. The dotted line
corresponds to the IBBM.
\newline(b) Same data than in (a) with a viscous dissipation 
proportional to the relative velocity between the grains and 
the lateral wall. The effective acceleration obtained from
the fit is $g_{\nu}=11.9\pm0.1\;m/s^2$.}

\label{fig.vueloviscoso}
\end{center}
\end{figure}

In order to improve the quality of the fitting,
we have tried another approach by introducing a viscous dissipation 
which depends on the velocity of the grains.
Then the ballistic trajectory of the grain will be 
modeled by the expression $\ddot{z}+\gamma \dot{z}=-g_{\nu}$ 
where $z$ is the position of the particle and $\gamma$ represents 
the dissipative coefficient.
The fit is shown in Fig ~\ref{fig.vueloviscoso}.a. The value
obtained for the initial velocity is in close agreement with the
experimental data, and the value for the effective
acceleration $g_{\nu}=12.77\pm0.07\;m/s^2$ is larger than the
obtained from the flight times. The dissipation can also be
taken into account by introducing a term which depends on the
relative velocity between the grains and the wall, so that
$\ddot{z}+b\dot{z} -g_{\nu} \left(\Gamma \; \mbox{sin}(\omega t + \phi)\; - 1 \right) =0$,
with $\dot z$ being now the relative velocity between the grains and the
lateral wall.
The fit is shown in Fig ~\ref{fig.vueloviscoso}.b. The value obtained
for the effective acceleration is now $g_{\nu}=11.9\pm0.1\;m/s^2$.
In both cases, the initial velocity is correctly reproduced.

It is not easy to choose which one of the fits is better. In any
case, the introduction of a dissipation which depends on the 
velocity clearly improves the quality of the analysis, and the
effective acceleration $g_{\nu}$ that is found for the particles
near the wall, where the trajectories of the grains have been
tracked, is larger than the one found from the flight 
times for all the granular layer as a whole, $g_{eff}$.
Therefore, the difference between $g_{eff}$ and $g_{\nu}$ must be 
due to the wall-particle interaction.
This difference between both values would amount to the
net stress that the walls cause on the granular layer, thus
imposing the downward flow near the wall.

A similar fit could be performed for the region after the period
doubling (Fig. ~\ref{fig.saltos}.b), but a better temporal
resolution is needed.

\section{Conclusions and discussion}

From the analysis carried out we can conclude the following.

For $\Gamma \lesssim 2.7$ the Inelastic Bouncing Ball Model reproduces
quite well the jumps of the granular medium as a whole, just
as had been previously reported for the dilation of the
upper layer \cite{vandoorn}. 
Nevertheless, flight times are a little bit
shorter than predicted by the model. We have verified that
the cause for this is the interstitial air. When the container
is evacuated, the agreement between the modified model and 
the measured data is excellent. We have therefore introduced
in the model the interaction between air and the grains so
that the agreement when there is air in the container is improved.

Above $\Gamma \sim 2.7$ the model is unable to faithfully reproduce 
the dynamics observed in the experiment. According to the model, the
flight time should increase monotonically until a saddle-node
bifurcation appears; then it should remain constant for a 
finite range of the control parameter. Afterwards it should suffer
a period-doubling bifurcation. The saddle-node bifurcation 
has not been observed in our experiments.
Besides, flight times are much lower than predicted both in the
presence of air and in vacuum. The finite duration of the 
collision between the granular layer and the vibrating plate
can be the cause for this discrepancy. A deeper study of this
subject will be presented elsewhere.

The critical value for the control parameter where the period
doubling bifurcation takes place is noticeably influenced by
the presence of interstitial gas. For a moderate vacuum, such
as the one reached in our experiment, the critical value of
$\Gamma$ approaches the predicted one for an inelastic model
but is still larger than it. 
Remarkably, flight times in vacuum below the bifurcation
are almost indistinguishable from those measured in air.
This suggests that the presence of air affects mainly the
stability of the solutions for the flight times after the
bifurcation.

Just above the period doubling bifurcation, the model fits
quite well the branches if the container is evacuated. In
air, the flight times are smaller than those predicted
by the IBBM. In neither case, however, the divergence predicted
by the model as $\Gamma$ grows larger is observed: flight 
times change slowly. This behaviour is probably caused by
the finite duration of the collision as well.

Therefore, even though near the points where the solutions
change (\emph{i. e.} $\Gamma=1$ and the period doubling
bifurcation) the system can be modeled as a perfectly
inelastic body, when the control parameter is increased
there is a certain point where the model loses validity.
Finite duration of the collision is suspected to be 
involved.

In this work we have focused in the region below the
bifurcation point, where flight times in air and in
vacuum are almost the same. From the study of the
trajectories described by the particles near the lateral
wall we have checked that they are subjected during
their flight to a net force larger than gravity.
A quadratic fit of the flight allows us to estimate the
value of effective gravity, that agrees quite well with 
the one inferred from the measurement of the flight
time. The quadratic fit implies a constant external
force.

The quadratic fit, however, does not yield a correct
value for the initial velocities (\emph{i. e.} the
velocities with which the grains take off when they
begin the flight). In order to improve this, we have
introduced a viscous dissipation. The
trajectory is correctly reproduced, including the
initial velocity. Remarkably, the value obtained
for the effective acceleration is bigger than that 
obtained from the flight time measurements. The difference 
between both effective accelerations
suggests that there is an extra force acting on the grains
near the walls larger than the average over all the granular
layer. This would be the cause of the downward movement
near the walls that sets the sense of the convective flow.

\section*{Acknowledgments}

This work has been funded by Spanish Government project
BFM2002-00414 and FIS2005-03881 as well as Acci\'on Integrada
HF2002-0015, by the local Government of Navarre, and by the
Universidad de Navarra (PIUNA). I.Z. and J.M.P. thank the
Asociaci\'on de Amigos de la Universidad de Navarra for a grant.
The authors wish to thank H. Kellay for his hospitality and his
comments and R. Ar\'evalo for his useful comments about the
numerical simulation.

\end{document}